\documentclass[pss,fleqn]{w-art}
\usepackage{times}
\usepackage{w-thm}
\usepackage[]{graphicx}
\begin{document}
\DOIsuffix{theDOIsuffix}
\Volume{3}
\Issue{11}
\Month{xx}
\Year{2006}
\pagespan{1}{}
\Receiveddate{1 May  2006}
\Reviseddate{}
\Accepteddate{14 June}
\Dateposted{24 November   2006}



\title[Short Title]{Mn-doped II-VI quantum dots:  artificial
 molecular magnets }


\author[Sh. First Author]{J. Fern\'andez-Rossier\footnote{Corresponding
     author: e-mail: {\sf jfrossier@ua.es}, Phone: +00\,999\,999\,999,
     Fax: +00\,999\,999\,999}\inst{1}} 
     \address[\inst{1}]{Departamento de F\`{\i}sica Aplicada, Universidad de
     Alicante, 03690 Spain}
\author[Sh. Second Author]{R. Aguado\inst{2}} 
     \address[\inst{2}]{Instituto de Ciencia de Materiales de Madrid, CSIC,
     28049 Madrid,Spain}

\begin{abstract}
The notion of artifical atom relies on the capability to change the
number of carriers one by one in  semiconductor quantum dots, and the resulting changes
in their electronic structure. Organic molecules with transition metal atoms
that have a net magnetic moment and display hysteretic behaviour are known
as single molecule magnets (SMM).  The fabrication of CdTe quantum dots 
chemically doped  with a controlled number of Mn atoms and with a number of 
carriers controlled either electrically or optically  paves the way towards
a new concept in nanomagnetism: the artificial single molecule magnet.  Here
we study the magnetic properties of a Mn-doped CdTe quantum dot for
different charge states and show to what extent they behave like a single
molecule magnet.

\end{abstract}
\maketitle                   

\section{Introduction}

One of the most studied single molecule magnets (SMM),  Mn$_{12}$, is made of a
cage of one hundred Carbon, Hydrogen and Oxygen atoms,  surrounding a  
Mn$_{12}$O$_{12}$ cluster \cite{Lis}. The oxydation state of the Mn atoms  
either $+3$  or $+4$ so that their spin is either (S=$2$) or (S=$3/2$). 
In the lowest energy configuration, the molecule has total spin $S=10$  that
results from  the   super-exchange interactions between the Mn spins
\cite{Sessoli}. The
$S=10$ manifold has 21 states  that, due to the spin-orbit interaction, are
split according to $E=-DS_z^2$, where $S_z$ is the spin projection along a
crystallographic axis $z$ \cite{Sessoli,Friedman96}. Remarkably, the magnetization of a molecular
crystal  of Mn$_{12}$ shows hysteresis, at temperatures larger than the
inter-molecular coupling, reflecting that each individual molecule behaves as a
magnet\cite{Friedman96}. In marked contrast to mono-domain magnets,   the
hysteresis curve of a Mn$_{12}$ crystal displays steps at well defined values of the applied field.
These are associated to quantum tunneling events of the magnetic moment.
Therefore,  the magnetic moment of SMM behaves quantum mechanically
\cite{Tejada}.

As an extension of the concept of artificial atom \cite{artificialatom},  we
propose the notion of artificial SMM:  a  II-VI semiconductor quantum dot,
chemically doped with Mn atoms and electrically doped with carriers so that 
they form a collective spin that has meta-stable states (hysteresis)
and  steps in the magnetization. As a
substituional impurity of the cation in the II-VI material \cite{Furdyna}, Mn
is in a $2+$ oxydation state with spin $S=5/2$. In our device there are a few
Mn atoms, sufficiently apart as to make the antiferromagnetic exchange
interaction between them completely negligible. The magnetic properties of the
dot are given by the controlled presence of electrons and/or holes in the
conduction and valence band quantum dot levels. Since both electrons and holes
are exchanged coupled to the Mn spin \cite{Furdyna}, indirect exchange
interactions couple the Mn spins to each other \cite{JFR04} in the dot. Even 
in the case of a dot doped with a  single Mn atom, we show that the anisotropic spin
interaction of the holes and the Mn
\cite{Besombes1,Besombes2,Besombes3,JFR05,JFR-RAS06,trion}  results in SMM
behaviour.

Most of the working principles  of this proposal  have been verified
experimentally. On one side,  the charge state of II-VI quantum dot can be
prepared, with a single electron accuracy, both by
electrical\cite{Klein-Nature,Forchel-APL,Alivisatos05} and optical means
\cite{Henneberger01,Besombes02}.   On the other side, several groups have reported the
fabrication if II-VI quantum dots doped with Mn atoms, and the exchange
coupling of electrons and holes to the Mn spins
\cite{Maksimov,SeufertPRL02,Bacher,Doro,Smith,Nature2005}. More recently, 
the controlled fabrication of CdTe quantum dots doped with a single Mn atom
\cite{Besombes1,Besombes2,Besombes3}, as well as  the electrical control of the
charge state of a single CdTe quantum dot doped with a single Mn atom
\cite{trion} and a GaAs island doped with tens of Mn atoms \cite{Wunderlich}
have been  reported. Electrical 
control of the carrier density has been demonstrated  in larger  magnetic
semiconductor structures, making it possible to alter reversibly properties
of the material like the $T_C$ \cite{Ohno2000,Boukari}    and the coercive
field \cite{Ohno2003} of these systems. Our proposal targets the achievement
of  this notion at the scale of a single electron and the single Mn atom
limit.

Related theory work  has focused on the electronic structure
\cite{JFR04,Climente,PRL2005-Haw,PRL2005-Abinit,Govorov05a}, 
optical response \cite{JFR05,Bhatta,Govorov04} and
transport \cite{JFR-RAS06,Efros} of Mn-doped quantum dots as a function of the
number of confined carriers. It is now well established that, depending
on the number of carriers present in the dot, the eigenstates of the
Mn-fermion Hamiltonian are very different \cite{JFR04}. In previous work
we have derived from numerical diagonalizations
\cite{JFR-RAS06}  the effective spin Hamiltonian  for
relevant  low energy sector of a CdTe dot doped with a single Mn interacting
with up to 3 carriers, either electrons or holes. Here we  give a
preliminary account of the magnetization curves  of a single CdTe dot, doped
with a Mn atom, as a function of the number of carriers.  The results
 presented here permit to claim that Mn-doped quantum dots with extra holes
behave like a nanomagnet, with hysteretic magnetization and steps of the
magnetization at precise values of the applied field.  

The rest of this paper is organized as follows. In section II we summarize the
model Hamiltonian for a Mn-doped CdTe quantum dot and  we review the properties
of the 
eigenstates and eigenvalues of the Hamiltonian for a given charge
state ${\cal Q}=\pm1,\pm2,\pm3$ .   In
section III the magnetization of the dot with ${\cal Q}=+1$ and a single
Mn is calculated. 
We first show that the equilibrium magnetization is anisotropic, in contrast to
the free Mn case, due to the interaction to the hole.   Then we discuss how the
magnetization  behaves if the field is varied so that equilibrium can not
be reached, which can be the standard situation provided the long spin
relaxation time of this system.

\section{Hamiltonian}

We consider quantum dots with a typical size of 10 nanometers and several
thousand atoms, beyond reach of present-day  ab-initio  calculations
\cite{PRL2005-Abinit}. This justifies the use of the  standard model
Hamiltonian
\cite{Furdyna,JFR04,JFR05,JFR-RAS06,Govorov05a,Bhatta,Govorov04,Efros} for
diluted magnetic semiconductors that describes   CB electrons and VB  holes 
interacting with localized Mn spins ($\vec{M}_I$) via a local exchange
interaction  and their coupling to an external magnetic field, $\vec{B}$. The Mn
spins interact also with each other via short range superexchange coupling.  In
general the Hamiltonian can not be solved exactly and  in most instances mean
field or some other approxiamations are used. Only in the limit of a few Mn
atoms and a few electron and hole states considered in this paper it is
possible to diagonalize the Hamiltonian numerically without approximations. The
second quantization Hamiltonian of the isolated dot reads:
\begin{eqnarray}
{\cal H}_{QD}=
\sum_{\alpha,\alpha'}
\left(
\epsilon_{\alpha}(\vec{B})
 \delta_{\alpha,\alpha'}
 +
\sum_I J_{\alpha,\alpha'}(I)
\vec{M}_I\cdot
\vec{S}_{\alpha,\alpha'}(\vec{R}_I)
\right)f^{\dagger}_{\alpha}f_{\alpha'}
+
 \sum_{I,J} J^{\rm AF}_{IJ}\vec{M}_I\cdot\vec{M}_J
\label{hamil}
\end{eqnarray}
Here $f^{\dagger}_{\alpha}$ creates  a band carrier in the
$\alpha$ single particle state of the quantum dot, which can be
either a valence band or a conduction band state. 
 QD carriers occupy localized  spin orbitals $\phi_{\alpha}$ with energy
$\epsilon_{\alpha}$ which are described in the envelope function
$\vec{k}\cdot\vec{p}$ approach \cite{JFR04,JFR05,JFR-RAS06,dot-holes}.
 In the case of
valence band holes the  6 band Kohn-Luttinger Hamiltonnian, including spin
orbit interaction, is used as a starting point to build the quantum dot
states \cite{dot-holes}.
The first term
in the Hamiltonian describes non interacting carriers in the dot
and the second term describes the exchange coupling of the
carriers and the Mn. We neglect  interband exchange so that
$J_{\alpha,\alpha'}=J_e$ ($J_{\alpha,\alpha'}=J_h$) if both
$\alpha$ and $\alpha'$ belong to the conduction band states
(valence band states). In contrast, we include 
exchange processes by which a
carrier is scattered between two different levels of the dot that
belong to the same band
 are included, although the level spacing
considered here is much larger than the exchange interaction. 
The matrix elements of both valence and conduction band spin density,
evaluated at the location of the Mn atoms, are given by
$\vec{S}_{\alpha,\alpha'}(\vec{R}_I)$. They
depend strongly on the orbital nature of the single particle level in question.
In the case of conduction band we neglect spin orbit interactions so
that $\vec{S}_{\alpha,\alpha'}$ is rotationally invariant
\cite{JFR04}. In contrast,  strong spin orbit interaction of the
valence band makes the Mn-hole interaction strongly anisotropic
\cite{JFR05,dot-holes,JFR01,TAMR} and it varies between different dot levels.
Following previous work \cite{JFR04,JFR05,dot-holes}
confinement is described by a hard wall cubic potential
 with   $L_z<L_x\simeq L_y$. Although  real dots are not cubic, 
 this simple model \cite{dot-holes} predicts properly that  
 the lowest energy hole doublet has a
strong heavy hole character, with spin quantized  along the growth
direction $z$ whereas the next single particle doublet is mainly
light hole with spin quantized in the $xy$ plane.

As discussed in our previous work \cite{JFR-RAS06}, in the case of ${\cal
Q}=\pm1$ and ${\cal
Q}=\pm3$ the low energy properties of
${\cal H}_{QD}$ are described by
\begin{equation}
{\cal H}_{eff}=
\sum_{I,a=x,y,z} j^a_I \tau^a M^a_I
+ 
 \sum_{a=x,y,z}  \mu_B B_a \left(g M^a_I + g_a({\cal Q}) \tau^a \right)
+
 \sum_{I,J} J^{\rm AF}_{IJ}\vec{M}_I\cdot\vec{M}_J
\label{effective}
\end{equation}
where $\tau^a$ are the Pauli matrices operating on the isospin space defined by
the lowest energy single particle doublet. In the case ${\cal Q}=-1,-3$ we have
$j^x=j^y=j^z$. In the case ${\cal Q}=+1$ we have $j^x=j^y=0$  in the absence of
light-heavy hole mixing ($L_x=L_y$). In the case ${\cal Q}=+3$ we have
$j^z<j^x=j^y$. 

Hereafter we only consider dots with one Mn atom. Therefore, Mn-Mn 
superexchange interaction is irrelevant. 
The coupling of Mn to conduction band electrons  is spin rotational invariant,
so that the ground state manifold for ${\cal Q}=-1,3$ are given by the 7 states
with $S=3$ and $S_z=\pm3, \pm2, \pm1,0$. In contrast, the Mn-hole coupling is
very anisotropic.  In a dot with ${\cal Q}=+1$ and $L_x=L_y$ spin-flip Mn-hole
interactions are strictly forbidden. The resulting  Ising coupling between the
Mn and the hole yields a spectrum with 6 doublets. Their wave functions are
given by $|M^z,S_h^z\rangle$ with
and the eigen-values $E_{M^z,S^z}=|j_z|M^z\times S^z $, where  $j^z=J_{h}
 |\psi(\vec{R}_{Mn})|^2$.  The lowest energy doublet corresponds to the Mn spin
 maximally polarized against the heavy hole spin, which lies along the growth
 direction, $z$.  The addition of a second  holes fills the heavy hole doublet.
 The low energy sector has 6 states,  spin splitted according the absolute
 value of the third component the Mn spin, the lowest energy is given by the
 $M_z=\pm 1/2$ doublet. This splitting   comes from inter-level exchange
 coupling, which are of order $\Delta^2/\delta$ \cite{JFR04}. . The addition of
 a third hole into the same dot leaves a doubly occupied heavy hole doublet and
 a singly occupied  light hole  interacting with the Mn. The different
 spin-orbital properties of the light hole result in an exchange coupling to
 the Mn which is in between the isotropic Heisenberg coupling of ${\cal Q}=-1$
 and the Ising coupling of ${\cal Q}=+1$. 
The picture emerging from these results is the following: the low energy states
of the Mn-fermion complex are spin-split at zero applied field. The spin
splitting $\Delta$ is much larger for open shell
configurations than for closed shell configurations. The symmetry properties of
the lowest energy states are different for electrons, heavy holes and light
holes.

\section{Magnetic response}
Now we address the magnetic properties of the single Mn doped CdTe dot although
the  results also scale for larger N. We calculate the many-body spectrum of
the dot as a function of an external magnetic field. Since the single particle
level spacing  is much
larger than the cyclotron frequency, it is not a bad approximation to  neglect
the orbital magnetism. The magnetization is given by :
\begin{equation}
\langle \vec{{\cal M}}\rangle = \sum_{N}  \langle N |\rho\left(
\sum_I g\mu_B\vec{M}_I + g_{f} \mu_B\vec{S}_f \right)|N\rangle=
\langle \vec{{\cal M}}\rangle_{Mn}+\langle \vec{{\cal M}}\rangle_{f}
\end{equation}
where $\rho$ is the quantum dot density matrix  and$g_{f}$ denotes the gyromagnetic
factor of electrons and holes.

\subsection{Equilibrium Results}
In a first stage we assume that the density matrix of the quantum dot is that
of thermal equilibrium. We consider the case of a dot with ${\cal Q}=+1$.
At zero magnetic field the eigenstates  with opposite magnetization are
degenerate so that their populations are identical and the average magnetization
vanishes. This is  a consequence of the ergodic approximation, that assigns
equal weight to the states of identical energy, and to time reversal symmetry.  
Application of a magnetic along the easy direction $z$  splits the energy levels
without mixing them, since $B_z M^z + B_z S^z$ conmutes with the Hamiltonian.
The resulting energy diagram is shown in figure 1, lower left panel, and
features numerous level crossings, reflecting the absence of transverse
interactions.  The magnetic field breaks the degeneracy of all the doublets,
including the  lowest energy one. 

At temperatures smaller than the Zeeman splitting only the ground state is
occupied, resulting in a net magnetization $\langle {\vec M} \rangle$, shown in
the upper panel of figure 1. We show the Mn  magnetization only because is
dominant and simplifies the discussion. In that figure we show the
magnetization of the dot with ${\cal Q}=0$, which is the standard Brillouin
function for $S=5/2$. It is apparent that the ${\cal Q}=0$ magnetization is
softer than the ${\cal Q}=+1$ magnetization curve when the field is applied
along the $z$ direction, due to the exchange splitting of the levels with
smaller $|M_z|$. This result is in agreement with previous work for exciton
polarons \cite{Bhatta95} In contrast, the magnetization curve when the field is
applied  in the $x$ direction, transverse to the easy axis of the dot, is much
smaller than the free case. This is due to the competition between exchange
interactions, that favor the alignment of the Mn along the $z$ axis and the
magnetic field. The difference is also seen in the energy spectrum, in the lower
right panel.   Hence,  the presence of a heavy hole in the dot would result in a
strong magnetic anisotropy in the equilibrium magnetization

\subsection{Non-equilibrium results}
The relaxation of the Mn-fermion spin towards equilibrium comes from the
release of energy and angular momentum to their environment\cite{Chud}. Since
this process can be very slow, we need to consider a non-equilibrium  density
matrix. The relaxation time    $T^{{\cal Q}=0}_1$ for Mn spins in very dilute
CdMnTe can be as long as $10^{-3}$ s\cite{PR60}. In the case of a $N={\cal
Q}=+1$ dot,  the transitions between the two members of the ground state, $|\pm
5/2,\mp 1/2\rangle$ require a energy conserving perturbation that 
changes the Mn spin by $5 \hbar$  and the hole angular momentum by 
$3\hbar$  in the opposite direction. Such a process is highly forbidden 
and his lifetime $T^{{\cal Q}=+1}_1$ will be
definitely larger than $T^{{\cal Q}=0}_1$ . Therefore, if the
applied field is varied more rapidly than $T^{{\cal Q}=+1}_1$ we are not
entitled to use the equilibrium density matrix.

Now we revisit the results of the previous subsection. We assume that
the Ising Hamiltonian is perturbed by very weak
additional terms that will only be effective close to degeneracy points in the
energy diagram of figure 2. They read:
\begin{equation}
{\cal V}= a \left((M^x)^4 +(M^y)^4\right) + 
\sum_{i} {\cal J}_i \vec{M}\cdot\vec{I}_i
+\epsilon \left(M^x \tau^x + M^y \tau^y\right)
\label{pert}
\end{equation} 
The so called cubic field term  is
the  lowest order magnetic anisotropy term
allowed by the crystal symmetry \cite{PR60} and makes it possible to flip the Mn
spin by 4 units. The second term comes from the hyperfine coupling of the Mn $d$
electrons to the nuclear spins $\vec{I}$ \cite{PR60}. It can flip the Mn spin by one unit. 
The last term
arises from light-hole heavy-hole mixing and permits the transition of the hole
via exchange of one unit of angular momentum with the Mn. Other terms like
dipolar coupling to the nuclear spins, of phonon-mediated hole spin relaxation
might be involved as well.

Our starting point is a the dot under the influence of a large magnetic field
along the positive $z$ axis at low temperatures so that  we assume that the dot
is in the ground state, $|+\frac{5}{2},\downarrow\rangle$. Now the applied
field is reduced down to zero, so that the state $E(+\frac{5}{2},\downarrow)$
becomes degenerate to $E(-\frac{5}{2},\uparrow)$.  None of the terms in
(\ref{pert}) can induce the transition from the former to the latter.
 It would take the combined action of
several  of them to achieve the spin-relaxing transition,
 which is thereby very slow.  Hence, it would
be possible to ramp the field through that degneracy point without spin
relaxation. In that case the magnetization at zero field would be finite. The
field could be reversed down to the crossing point of the state
$|+\frac{5}{2},\downarrow\rangle$ with the state
$|-\frac{3}{2},\uparrow\rangle$.  The cubic field term can flip the Mn spin,
and it would take further spin relaxation of the hole to achieve that
transition. The quantum  transition could take place here, or it could happen
at larger fields, where the initial state is
brought in to coincidence with $|-\frac{3}{2},\downarrow\rangle$, which is
accesible via  the cubic field term only. Further increase of
the field towards negative value would bring into resonance the 
$|-\frac{3}{2},\sigma_h\rangle$ levels  with the 
$|-\frac{5}{2},\uparrow\rangle$, completing the magnetization reversal
 process. In any  case the
magnetization  displays remanence,  hysteresis and abrupt steps  as we
show in the lower panel of figure 2. These are very similar  to
those observed in single molecule magnets \cite{Sessoli,Friedman96,Tejada}. 
This phenomenology is the consequence of 
the blocking of magnetization relaxation  processes except at 
 well defined values of the applied field that put
  levels with different $M_z$ in resonance. 

\section{Discussion and Conclusions}
Artifical atoms mimic the properties of natural atoms at much larger (smaller)
length (energy) scale. From that point of view,  artificial SMM is a hybrid
device.  The magnetic moment of both artificial and standard SMM comes from the
Mn atoms. What makes them different is the glue that determines the total 
magnetic moment of the SMM.    The properties of Mn$_{12}$ come from the
chemical bonding between the Mn and the oxygen atoms in the central core of the
molecule. In the case of artificial SMM, the glue electrons are delocalized in
a much larger length scale.

Measurement of the quantum features in the magnetization of molecular
magnets is greatly simplified due to the homogeneity of the single molecule
magnets in macroscopic molecular crystals.   In the case of artificial molecular
magnets, these specific values of the field that result in steps in the
magnetization depend on
the Mn-hole exchange coupling constant, which  depends both on the dot size
and on the Mn location inside the dot.  Therefore, the measurement of
magnetization over an ensemble of dots would yield results different from those
shown in figure 3. 

The measurement of the  effect predicted here would require either the
fabrication of extremely mono-disperse dots, using for instance 
colloidal techniques \cite{Gamelin1}, or the measurement of the magnetization of
a single device. The second possibility is starting to be explored
experimentally  in
the case of  Mn$_{12}$ molecules \cite{CM1,CM2}.  We have
studied recently  the theory  of a
similar device in which Mn$_{12}$ is replaced by a single Mn-doped CdTe quantum
dot . We have shown that the magnetic state of the Mn
determines the conductance of the dot. This opens the door for the measurement
of $M(B)$ in a single dot. An additional experimental route could come from
spin-selective  optical  measurements of a single dot.   
In summary, we claim that a Mn-doped CdTe quantum dot charged with a single
heavy hole will have exotic magnetic properties that correspond to those of
single molecule magnets.  A single electron transistor with a single-Mn doped
CdTe quantum dot would behave like an artificial single molecule magnet.

\begin{acknowledgement}
This work
has been financially supported by MEC-Spain (Grants FIS200402356,
MAT2005-07369-C03-03, and the Ramon y Cajal Program) and by
Generalitat Valenciana (GV05-152).  This work has been partly
funded by FEDER funds.
\end{acknowledgement}
\begin{figure}[thb]
\begin{minipage}[t]{.45\textwidth}
\includegraphics[width=\textwidth]{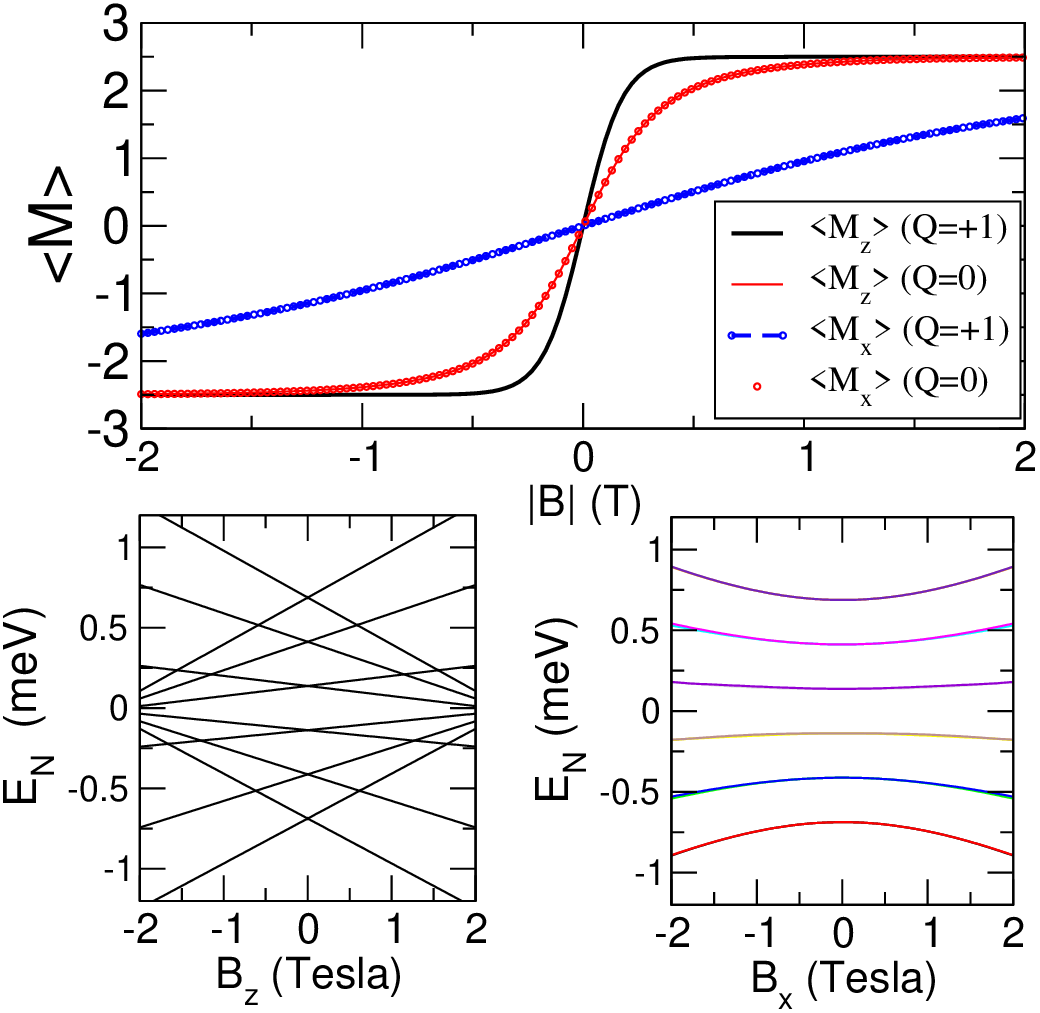}
\caption{Upper panel: equilibrium magnetization for dot with ${\cal Q}=0$ and
${\cal Q}=+1$ with the field applied both along the growth axis and in-plane.
Lower Panel: Energy spectrum for the dot as a function of $B_z$ (left) and $B_x$
(right). }
\label{fig:3}
\end{minipage}
\hfil
\begin{minipage}[t]{.45\textwidth}
\includegraphics[width=2.0in]{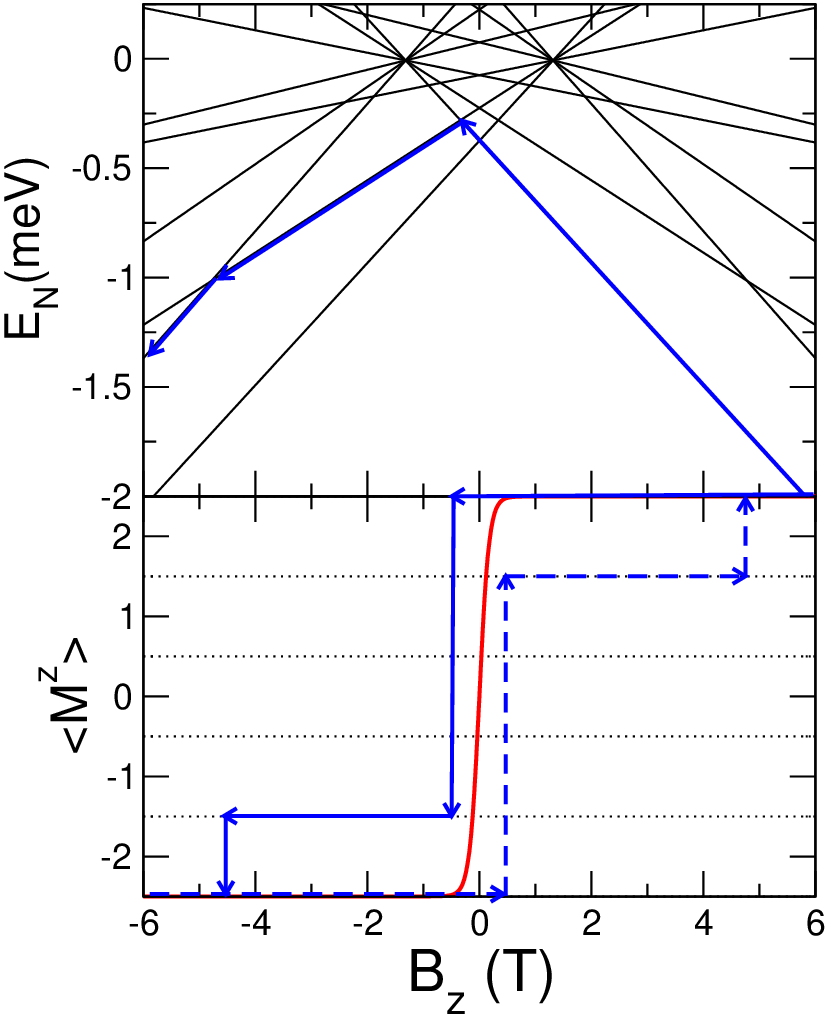}
\caption{Upper panel: detail of the Mn-hole spectrum and a possible
non-ergodic evolution of the density matrix. 
Lower panel: non-ergodic magnetization  of the system.}
\label{fig:4}
\end{minipage}
\end{figure}

\end{document}